\documentclass[12pt, a4paper]{article}

\usepackage[utf8]{inputenc}
\usepackage{amsmath, nccmath}
\usepackage{float}
\usepackage{graphicx}
\usepackage{geometry}
\usepackage[hidelinks]{hyperref}
\usepackage{caption}
\usepackage{lmodern}
\usepackage{amsfonts}
\usepackage{fancyhdr}
\usepackage{parskip}
\usepackage[style=numeric-comp, sorting=none]{biblatex}
\usepackage{algorithm}      
\usepackage{algpseudocode}

\begin{document}
\thispagestyle{fancy} 

\begin{center}

    \hrule
    \vspace{1.5ex}
    {\Large \bfseries A Minimal Quantitative Model of Perceptual Suppression and Breakthrough in Visual Rivalry \par}
    \vspace{1.5ex}
    \hrule
    \vspace{3em}

    {
      \begin{tabular}{c}
      Christopher J. Whyte\textsuperscript{1,2,*},
      Hugh R. Wilson\textsuperscript{3},
      Shay Tobin\textsuperscript{4},
      Brandon R. Munn\textsuperscript{1,2}, \\
      Shervin Safavi\textsuperscript{5,6},
      Eli J. Muller\textsuperscript{1,2},
      Jayson Jeganathan\textsuperscript{1,2},
      Matt Davidson\textsuperscript{7,8}, \\
      James M. Shine\textsuperscript{1,2,$\dagger$}, and
      David Alais\textsuperscript{7,$\dagger$}
      \end{tabular}
    \par}
    \vspace{2em}

    {\small
      \textsuperscript{1}Centre for Complex Systems, University of Sydney, Sydney, Australia\\
      \textsuperscript{2}Neuroscience Research Theme, School of Medical Sciences, Faculty of Medicine and Health, University of Sydney, Sydney, Australia\\
      \textsuperscript{3}Centre for Vision Research, York University, Toronto, Canada\\
      \textsuperscript{4}School of Mathematical and Physical Sciences, Macquarie University, Australia\\
      \textsuperscript{5}Computational
        Neuroscience, Department of Child and Adolescent Psychiatry, Faculty of Medicine, Technische Universität Dresden, Dresden, Germany\\

      \textsuperscript{6}Department of Computational Neuroscience, Max Planck Institute for Biological Cybernetics, Tübingen, Germany\\
      \textsuperscript{7}Discipline of Psychology, Graduate School of Health, University of Technology Sydney, Sydney, Australia\\
      \textsuperscript{8}School of Psychology, University of Sydney, Sydney, Australia\\[1ex]
      \textsuperscript{*}\textit{Corresponding author: christopherjackwhyte@gmail.com}\\
      \textsuperscript{$\dagger$}\textit{Co-senior author}
    \par}
    \vspace{3em}

\end{center}

\begin{abstract}

\noindent When conflicting images are presented to either eye binocular fusion is disrupted. Rather than experiencing a blend of both percepts, often only one eye’s image is experienced, whilst the other is suppressed from awareness. Importantly, suppression is transient – the two rival images compete for dominance, with stochastic switches between mutually exclusive percepts occurring every few seconds with law-like regularity. From the perspective of dynamical systems theory, visual rivalry offers an experimentally tractable window into the dynamical mechanisms governing perceptual awareness. In a recently developed visual rivalry paradigm – tracking continuous flash suppression (tCFS) – it was shown that the transition between awareness and suppression is hysteretic, with a higher contrast threshold required for a stimulus to breakthrough suppression into awareness than to be suppressed from awareness. Here, we present an analytically-tractable model of visual rivalry that quantitatively explains the hysteretic transition between periods of awareness and suppression in tCFS. Grounded in the theory of neural dynamics, we derive closed-form expressions for the duration of perceptual dominance and suppression, and for the degree of hysteresis (i.e., the depth of perceptual suppression), as a function of model parameters. Finally, our model yields a series of novel behavioural predictions, the first of which – distributions of dominance and suppression durations during tCFS should be approximately equal - we empirically validate in human psychophysical data.
\end{abstract}

\clearpage

\section{Introduction}

When incompatible stimuli are presented to each eye, binocular fusion is prevented and the stimulus presented to one eye is consciously experienced, whilst the other is suppressed from awareness \cite{Blake1989, Blake2001, Blake2002, Blake2022}. Suppression is, however, transitory, and the two rival stimuli stochastically enter and disappear from awareness with a characteristic right-skewed distribution of dominance durations \cite{Fox1967, Levelt1967}. Visual rivalry resulting from interocular conflict, therefore, offers a psychophysically tractable window into the dynamical mechanisms underlying the transition between perceptual awareness and suppression, attracting considerable attention from experimentalists \cite{Logothetis1996, Mitchell2004, Alais2007, Blake1979, Hollins1981, Lunghi2015, Nguyen2003, Teng1994, Tsuchiya2006, Gelbard2018, hesse2020new, Kapoor2022, Drew2022, Dwarakanath2023, Montgomery2025}, and neural modellers \cite{Dayan1998, Theodoni2011, Laing2002, Moreno-Bote2007, Safavi2024, Shpiro2007, whyte2025burst, Wilson2003, Wilson2007, Gershman2012, Wilson2017, Li2017, Leptourgos2020, Cao2021, Barkdoll2023, Wu2025} alike.

In visual rivalry paradigms that utilise stimuli with constant contrast, the minimal physiological conditions for awareness and suppression are theoretically well understood \cite{Seely2011, Shpiro2007, Wilson2007}. In binocular rivalry (the simplest and most common form of interoccular rivalry), two conflicting static stimuli are presented to each eye, and perception alternates between the two mutually exclusive stimulus percepts \cite{Blake2001}. Here, adaptation and intrinsic noise both play an essential role \cite{Brascamp2006, Shpiro2009, Braun2010, Theodoni2011, Pastukhov2013, Cao2021, Wu2025}. Models poised at the boundary between noisy excursions around a stable adaptation-driven limit cycle, and noise-driven jumps between adaptation-modulated fixpoint attractors provide the best fit to empirical data \cite{Shpiro2009}.

The relationship between the physical properties of the stimuli entering each eye and the duration of the resulting mutually exclusive rival percepts is remarkably consistent and can be captured by a small set of law-like propositions known as Levelt’s laws \cite{Levelt1965}. The propositions have proven extremely robust, receiving only minor modification in the 60 years since their conception \cite{Brascamp2015}. Levelt’s laws have, therefore, served as an explanatory benchmark for computational models of binocular rivalry. As a result, the dynamical basis of the laws is, likewise, theoretically well understood. Closed-form expressions for each law can be derived from models with physiologically meaningful parameters \cite{Laing2002, Seely2011, Wilson2007}, offering a quantitative explanation of the relative dominance of each percept in terms of the underlying dynamical processes and their coarse-grained physiological implementation.

The minimal conditions for awareness and suppression in continuous flash suppression (CFS;\cite{Tsuchiya2005}), another canonical rivalry paradigm, can also be understood in terms of the same underlying principles. In CFS, a mask consisting of a sequence of high-contrast random patterns is presented to one eye and a static target stimulus is presented to the other. Unlike binocular rivalry where stochastic perceptual alternations occur every few seconds, CFS typically entails a single asymmetric period of dominance, whereby the dynamic random pattern of the mask is perceived for tens of seconds. Observers typically report the moment the weaker, previously suppressed visual target breaks free from suppression and enters awareness. Importantly, the prolonged period of mask dominance is thought to be enabled through a reduction in adaptation. Indeed, Shimaoka and Kaneko \cite{Shimaoka2011} showed that modifying a model of binocular rivalry so that the mask periodically drives different orientation selective populations leads to lengthened periods of suppression, on the order of tens of seconds, consistent with CFS.

The central appeal of visual rivalry paradigms is that although two visual stimuli enter the visual system, only one is consciously experienced whilst the other is suppressed, offering a practical way to experimentally study the mechanisms underlying visual awareness and suppression. However, in nearly all CFS studies, the threshold for awareness and suppression is confounded with the accumulation of adaptation. Although it is possible to study the dynamics of perceptual awareness and suppression, neither paradigm offers a reliable or practical means of measuring the threshold for awareness or suppression.

To overcome this limitation Alais and colleagues \cite{Alais2024} introduced a novel variant of CFS known as tracking continuous flash suppression (tCFS) which captures periods of suppression and awareness during continuous flash suppression. In tCFS, a high contrast dynamic mask is presented to one eye, whilst the contrast of the target stimulus presented to the other eye changes (log) linearly as a function of the observer’s perceptual report. The target stimulus starts at high contrast, dominating perception, and steadily decreases in contrast until the observer reports that it has been suppressed from awareness by the mask. Upon report the contrast of the target stimulus reverses direction and increase until it breaks through suppression into awareness (continuing in a cycle). Stimulus contrasts registered at points of suppression and breakthrough provide threshold estimates for both breakthrough (i.e. awareness) and suppression. Importantly, manipulating the contrast rate of the target stimulus provides a tractable way to manipulate the contribution of adaptation, with faster contrast rates giving less time for adaptation in the suppressed population to decay before it becomes dominant again. Intriguingly, stimulus contrast thresholds measured at points of breakthrough and suppression show a robust pattern of hysteresis (known as suppression depth in the visual rivalry literature), with a higher contrast threshold required for a stimulus to breakthrough suppression into awareness, than to be suppressed from awareness. In addition, the depth of hysteresis was shown to increase as a function of contrast rate, pointing to a systematic role for adaptation in setting the threshold for awareness and suppression. tCFS thus offers an opportunity to systematically examine the role of adaptation in setting the threshold for perceptual breakthrough and suppression.

Here we develop a theoretical account of the neural mechanisms that govern awareness and suppression in tCFS. In a companion paper \cite{whyte2025minimal}, we generalised a minimal model of binocular rivalry \cite{Wilson2007} to tCFS, and used numerical simulations to show that a unified mechanism, competitive inhibition between slowly adapting monocular populations, can account for perceptual awareness and suppression in both paradigms. Here, we use the same model to develop a quantitative account of the mechanisms governing awareness and suppression. Specifically, we extend the quantitative study of visual rivalry to the threshold for awareness and suppression by deriving closed-form expressions for dominance and suppression durations, and for the degree of hysteresis separating breakthrough and suppression thresholds, as a function of model parameters. We then leverage the expression for the depth of hysteresis to propose a novel empirical prediction, which we confirm in previously collected psychophysical data \cite{Alais2024}. Finally, we derive two additional behavioural predictions that can be tested in future experiments.

\section{Model}

\begin{figure}[h!]
    \centering
    \includegraphics[width=1\textwidth]{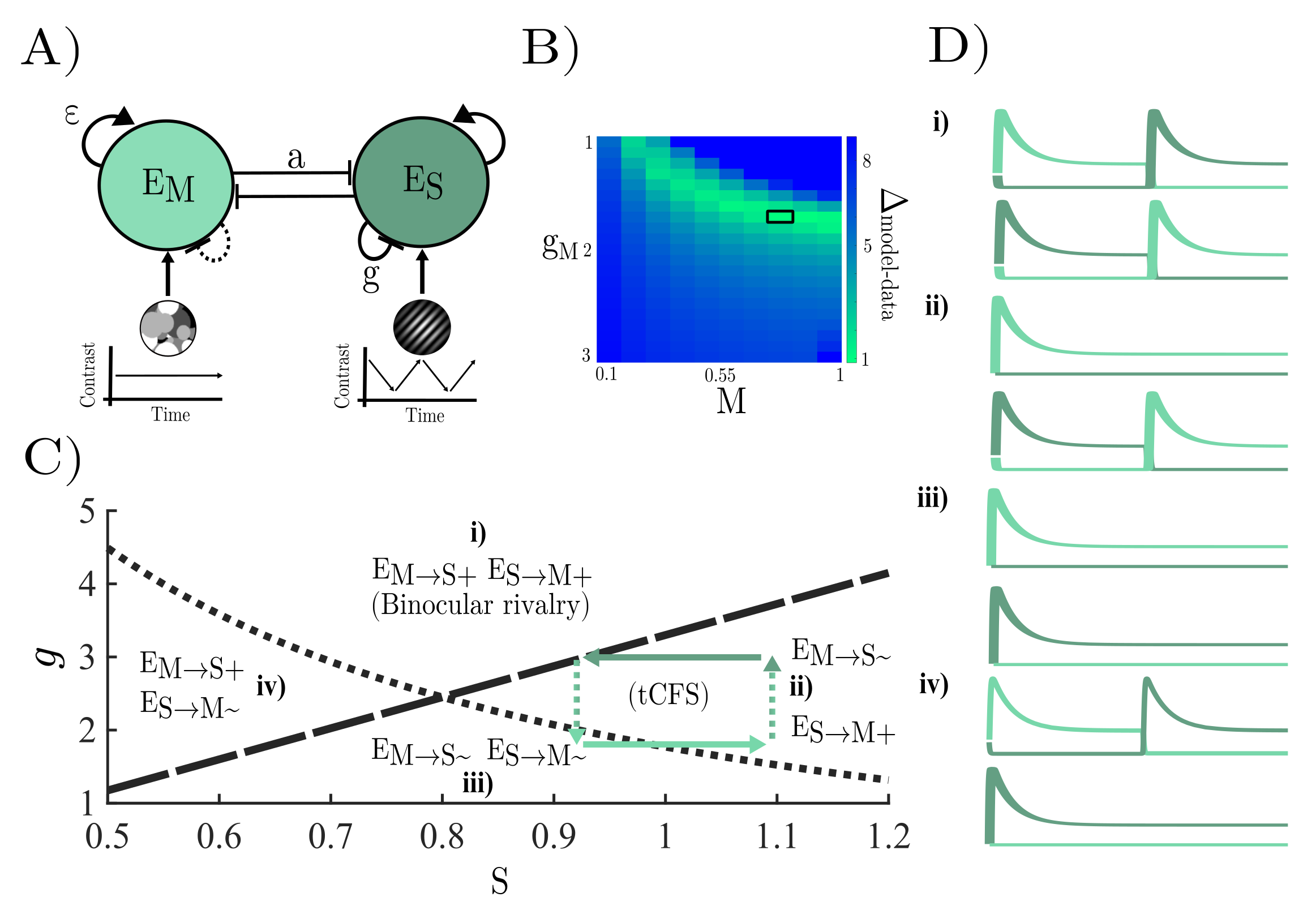}
    \caption{\footnotesize{\textbf{A)} tCFS model architecture – monocular populations driven by a mask (M) with constant contrast, and target stimulus (S) with percept-dependent ramping contrast, compete through mutual inhibition with asymptotic dynamics governed by adaptation. \textbf{B)} Dominance duration loss $(\Delta_{\text{model-data}})$ quantifying the fit between empirical and simulated dominance durations across the $[g_M \times M]$ parameter space. The best fitting parameter combination $g_M=1.7, M=0.8$ is highlighted by the solid black rectangle. \textbf{C)} The model $[g \times S]$ parameter space is partitioned into four distinct dynamical regimes by the inequalities in equation 6 ($g_M$- dotted line, $g_S$- dashed line): regime i) both populations escape suppression asymptotically via adaptation; regime ii) if the mask population is dominant, and the stimulus is suppressed, adaptation alone will not result in a perceptual switch, but if the stimulus is dominant the mask will asymptotically escape suppression via adaptation; regime iii) both the mask and target stimulus population cannot escape from suppression by adaptation alone; regime iv) if the stimulus population is dominant and the mask population is suppressed, adaptation alone will not result in a perceptual switch, but if the mask is dominant, the stimulus population will escape suppression via adaptation. An example trajectory across a full tCFS dominance-suppression cycle is shown in the green arrows. The dark green arrow represents periods where the stimulus-driven population is dominant. The stimulus starts at a high contrast value and gradually decreases in contrast until the mask-driven population (light green) escapes from suppression and becomes dominant, at which point the contrast of the stimulus starts to increase until the population driven by the stimulus escapes from suppression (and so on). \textbf{D)} Stylised representation of each dynamical regime.}}
    \label{figure:1}
\end{figure}

The dynamics of the model are described by a system of four nonlinear ordinary differential equations (ODEs) with: reciprocal inhibition between competing monocular populations; self-adaptation; and percept-dependent stimulus contrast ramping (Figure \ref{figure:1}A). This combination guarantees alternations between mutually-exclusive perceptual states,
\begin{align}
\tau_E \dot{E}_M &= -E_M + f(M + \varepsilon E_M - a E_S - g_M H_M)  \label{eq:1}  \\
\tau_H \dot{H}_M &= -H_M + E_M \nonumber \\
\tau_E \dot{E}_S &= -E_S + f(S + \varepsilon E_S - a E_M - g_S H_S) \nonumber \\
\tau_H \dot{H}_S &= -H_S + E_S\,, \nonumber
\end{align}
where the four state variables – $E_M, E_S$ and $H_M, H_S$ – represent the aggregate excitatory neural activity of orientation-selective monocular populations driven by the mask and stimulus (denoted by the subscripts $M$ and $S$ respectively), and the aggregate slow hyperpolarising adaptation current of each population. For analytic tractability, the transfer function is a simple threshold nonlinearity $f(x) = \max(x,0)$. The parameters $\varepsilon, a,$ and $g$, are dimensionless and represent the strength of recurrent excitation, competitive inhibition, and adaptation, respectively. The population time constants $\tau_E$ and $\tau_H$ are in units of milliseconds. Each neuronal population receives an independent (dimensionless) external drive ($M , S$). Unlike binocular rivalry, where the external drive shown to both eyes is stationary, in tCFS, a dynamic (flashing) mask is presented to one eye and the contrast of the target stimulus presented to the other eye changes (log) linearly with time. To approximate the effect of the dynamic mask whilst keeping the model analytically tractable, we drove the mask population with a constant stimulus but reduced the strength of adaptation in line with previous hypotheses \cite{Alais2024} and modelling \cite{Shimaoka2011}. Model parameter values and numerical simulation details are supplied in Appendix 1 and (unless stated otherwise) are the same as those used by Wilson \cite{Wilson2007}. To model the percept-dependent cyclic linear increase and decrease in stimulus contrast, we represented the stimulus with a piecewise linear ODE with dynamics that depend on which population is dominant:
\begin{equation}
\dot{S} = \begin{cases} -\gamma & \text{if } E_S > E_M \\ \gamma & \text{if } E_S < E_M\,, \end{cases}
\label{eq:2}
\end{equation}
where $\gamma$ is a dimensionless rate constant. We chose values of $\gamma$ so that the difference between the maximum and minimum contrast rate values used by Alais and colleagues \cite{Alais2024} was preserved. We matched $\gamma$ to empirical data by converting the empirical contrast rates (originally in units of dB per frame with a refresh rate of 60Hz) to units of dB per millisecond and used the difference between the maximum and minimum contrast rates (0.0021 – 0.0063 dB/ms) to define the range over which $\gamma$ varied. We then multiplied the full range of $\gamma$ values by a constant on the interval ($10^{0}-10^{-4}$), and found the value ($10^{-2}$) that resulted in simulated dominance durations with the same order of magnitude as the empirical data.

Instead of analysing the dynamics of the full, four-dimensional nonlinear system, we took advantage of the separation of timescales between the (fast) neuronal and (slow) adaptation variables (i.e. $\tau_H \gg \tau_E$; \cite{Gerstner2014, Wang2025}). This allowed us to partition the dynamics into two phases, both of which could be studied on the phase plane. In the first instantaneous phase $H_L, H_R \approx 0$, and in the second asymptotic phase, $H_L \approx E_L$ and $H_R \approx E_R$. In the initial instantaneous phase of the dynamics, if the argument of $f(x)$ is positive, equation \ref{eq:1} can be reduced to a two-dimensional system with Jacobian matrix:
\begin{equation}
J = \frac{1}{\tau_E} \begin{bmatrix} -1+\varepsilon & -a \\ -a & -1+\varepsilon \end{bmatrix}\,.
\label{eq:3}
\end{equation}
Geometrically, the existence of mutually exclusive perceptual states requires that the two asymptotically stable (mutually exclusive) states (corresponding to $E_M > 0, E_S=0$ and $E_M=0$ and $E_S>0$) are separated by a saddle point. That is, the eigenvalues of $J$ must have different signs. The eigenvalues are $\lambda_1 = \tau_E^{-1}(-1-a+\varepsilon)$, and $\lambda_2 = \tau_E^{-1}(-1+\varepsilon+a)$ for all physiologically plausible parameter values $a, \varepsilon > 0$ and $a > \varepsilon$, and as such $\lambda_1 < 0$, leading to the following inequality for $\lambda_2>0$:
\begin{equation}
a > 1-\varepsilon\,.
\label{eq:4}
\end{equation}
If the inequality in equation \ref{eq:4} is satisfied, the model will converge to one of the two winner-take-all states. As adaptation dynamics never reduce neural activity below the rectification threshold of the transfer function, once the system is in a winner-take-all state, the state will be stable as long as the competing population is suppressed. Perceptual switches occur when the competing population escapes suppression \cite{Brascamp2006, Shpiro2009, Braun2010, Theodoni2011, Pastukhov2013}. For example, if the mask-driven population $E_M$ is dominant, and the stimulus-driven population $E_S$ is suppressed, a perceptual switch will occur when $E_S$ escapes suppression and inhibits $E_M$, which, for the simple threshold nonlinearity, occurs when the argument of the population transfer function $f(x)$ passes through zero from below. We refer to the argument of $f(x)$ as aggregate synaptic drive $D$:
\begin{align}
D_M &= M + \varepsilon E_M(t) - a E_S(t) - g_M H_M(t), \label{eq:5} \\ 
D_S &= S + \varepsilon E_S(t) - a E_M(t) - g_S H_S(t)\,, \nonumber
\end{align}
where $D$ takes negative values when the population is suppressed and positive values when dominant. There are two types of perceptual switches that can occur within the model: adaptation-driven switches and stimulus-driven switches. Adaptation-driven switches occur when $g_{M/S}$ is strong enough to reduce the asymptotic firing rate of the dominant population to such a degree that the suppressed population is able to escape from inhibition. Such dynamics are likely solely governed by local circuit interactions. In contrast, input-driven switches occur when the self-inhibitory effect of adaptation is not sufficient for the suppressed population to escape from inhibition, and an increase in external drive to the suppressed population or a decrease in external drive (or possibly an increase/decrease in input from another cortical or subcortical region \cite{Gelbard2018}) to the dominant population is necessary for a switch to occur. We can find the value of $g$ separating these two regimes for each population by demanding that $D>0$ and solving for the value of $g_{M/S}$ necessary for the suppressed population to escape from inhibition. For the asymptotic escape of the stimulus-driven population ($E_S$) from suppression by the mask-driven population ($E_M$), we obtain the following inequality:
\begin{equation}
g_M > a\frac{M}{S} - 1 + \varepsilon\,,
\label{eq:6}
\end{equation}
where we have relied on the fact that at a perceptual switch the model is in the second asymptotic phase of its dynamics, where $H_S \approx 0$, and $H_M \approx E_M$, and when $E_S$ is suppressed $\varepsilon E_S=0$. We find the equilibrium value of the (dominant) mask-driven population by substituting $H_M = E_M$ into the equation for $\dot{E}_M=0$, and solve for $E_M$ leading to $E_{M}(\infty) = M/(1+g_M-\varepsilon)$, which when substituted into the expression for $D_S$ in equation \ref{eq:5} gives the inequality in equation \ref{eq:6} assuming that adaptation in the suppressed population has sufficiently decayed (close to zero) before the switch occurs. The comparable expression for the asymptotic escape of the mask-driven population is $g_S > a\frac{S}{M}- 1 + \varepsilon$.

This leaves us with two inequalities defining the boundaries of four distinct dynamical regimes (i.e., switch type × monocular population) in the model's parameter space (Figure \ref{figure:1}C). As suppression durations in CFS ($\sim$10 – 50 s) are much greater than the timescale of the adaptation current ($\sim$1 s), we hypothesised that the dynamics of the model that best represent tCFS would be located in the stimulus driven switching regime when the mask driven population is dominant (corresponding to regimes ii or iii in Figure \ref{figure:1}C), and be in the adaptation-driven switching regime when the stimulus driven population is dominant (i.e., regimes i or ii in Figure \ref{figure:1}C). As we did not want to simply assume this to be the case, we constrained adaptation and external drive values for the mask-driven population with behavioural data from Alais and colleagues \cite{Alais2024}.

Following Wilson \cite{Wilson2007}, we fixed the adaptation current parameter for the stimulus-driven population by finding the value of $g_S$ that resulted in an asymptotic firing rate that is approximately 30\% of the maximum firing rate at stimulus onset, as is observed empirically \cite{McCormick1989}. We used this value as the upper bound on the strength of adaptation for the mask-driven population which we fit empirically. Specifically, we ran a grid search over the $g_M \times M$ parameter space to find the parameter value combination that best minimised the difference ($\Delta_{\text{model-data}}$) between simulated and empirical dominance durations across slow (0.0021 dB/ms), intermediate (0.0042 dB/ms), and fast (0.0063 dB/ms) contrast rates (Figure \ref{figure:1}B). In the model, these stimulus values correspond to $\gamma=0.0021, 0.0042, 0.0063 \times 10^{-2}$. The best fitting parameter combination was $g_M=1.7, M=0.8$. In line with our hypothesis, the trajectory of the best fitting parameters remained within regimes ii) and iii) (Figure \ref{figure:1}C). All results reported below were stable across parameter combinations adjacent to the minima of $\Delta_{\text{model-data}}$.

\section{Hysteresis and contrast rate}

\begin{figure}[h!]
    \centering
    \includegraphics[width=.8\textwidth]{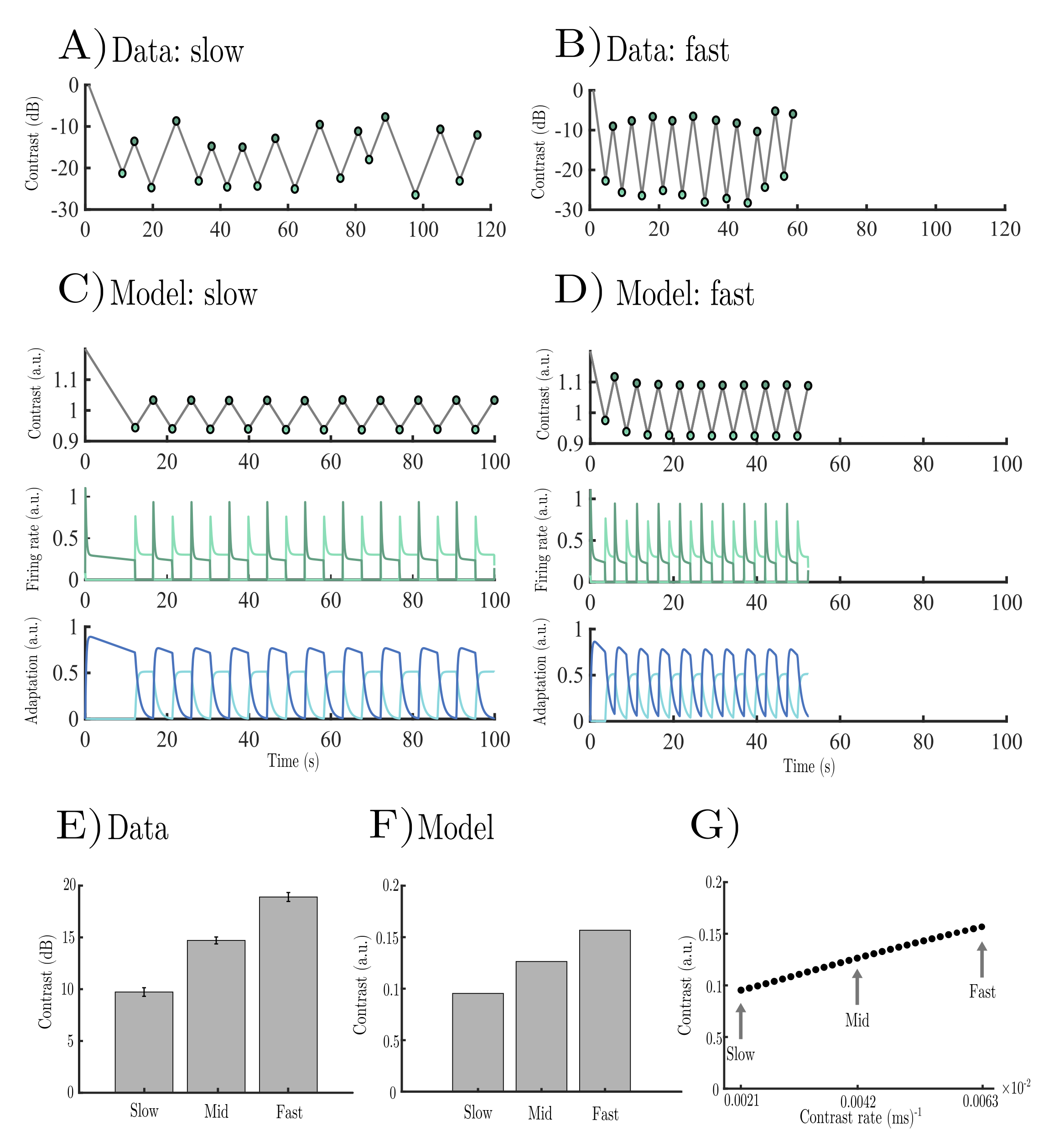}
    \caption{\footnotesize{\textbf{A)} Stimulus contrast dynamics from two example tCFS trials with slow and fast contrast rates adapted from \cite{Alais2024}. Points of suppression where the mask becomes dominant are highlighted with light green dots and points of perceptual breakthrough are highlighted with dark green dots. Each trial terminated after 20 reversals. \textbf{B-C)} Example simulated contrast dynamics, firing rates, and adaptation current dynamics for the population driven by the mask (light green) and the target stimulus (dark green). To aid in comparison with empirical data only 20 reversals are shown. We note that this is for purposes of illustration only, all simulated trials lasted for 120s. \textbf{D)} Approximate bifurcation diagrams with example model trajectories for the population driven by the target stimulus (upper dark green), and the population driven by the mask (lower light green). \textbf{E)} Empirical data for hysteresis depth across slow, mid, and fast contrast rates adapted from Alais and colleagues \cite{Alais2024}. \textbf{F)} Simulated hysteresis depth for matched slow, mid, and fast contrast rates. \textbf{G)} Hysteresis depth across the full range of contrast rates used in simulation. }}
    \label{figure:2}
\end{figure}

In tracking continuous flash suppression (tCFS; \cite{Alais2024}), the target stimulus is presented to one eye, and the flashing mask with constant contrast is presented to the other. The target starts at full contrast, and decreases log linearly (in units of dB) until the target stimulus is suppressed from awareness and the mask dominates perception. This perceptual reversal is indicated by a button press at which point the direction of contrast change reverses, continuing in a cycle until the trial is terminated (typical trials consist of 16 - 20 perceptual reversals; Figure \ref{figure:2}A-B). 

Two key empirical findings emerge from this paradigm. First, the contrast threshold for breakthrough and suppression is hysteretic. Across three experiments with multiple stimulus categories, the average difference between breakthrough and suppression thresholds was shown to be constant, with a reliably higher threshold for breakthrough than suppression. That is, the same stimulus, with the same contrast could dominant the content of awareness, or be suppressed from awareness, depending on the participant’s history of perceptual stimulation. Second, the degree of hysteresis is dependent on the rate of contrast change, with faster rates of contrast change leading to larger differences in breakthrough and suppression thresholds. This increasing difference in thresholds as a function of contrast rate points to a role for adaptation in accounting for the depth of hysteresis (Figure 2A-B and 2E; \cite{Alais2024}).

To understand these results mechanistically, we simulated tCFS across 30 contrast rates evenly distributed across an interval proportional to the difference between the maximum and minimum contrast rates that were used empirically \cite{Alais2024}. Each trial started by driving the model with a mask (with constant contrast) and an initially high contrast target stimulus. As was seen empirically, the stimulus-selective population won the initial competition for dominance, and remained dominant until the contrast of the target stimulus decreased sufficiently for the mask-driven population to escape suppression and inhibit the stimulus-selective population into silence (Figure \ref{figure:2}C-D). At this point, the target stimulus started to increase in contrast again until the external drive entering the stimulus selective population was sufficient for the population to breakthrough suppression from the mask. Across all contrast rates, the model converged to an equilibrium with stationary breakthrough and suppression thresholds. Initial differences between successive threshold values, due to the target stimulus starting with high contrast, decayed away as adaptation approached its equilibrium value across successive cycles of breakthrough and suppression.

In line with empirical data, in our simulations, the depth of hysteresis in contrast thresholds increased as a linear function of contrast rate. With our analytical approach, we can explain this trend quantitatively by deriving expressions for the value of the target stimulus at breakthrough ($S_B$) and suppression ($S_S$) points. At the point of perceptual breakthrough, the dominant mask-driven population is well approximated by its asymptotic value $E_{M}(S_B) \approx E_{M}(\infty) = M/(1+g_M-\varepsilon)$, and the stimulus-driven population is suppressed $E_S=0$. Substituting these values into equation \ref{eq:5} and setting $D_S=0$, we can then solve for $S_B$ leading to:
\begin{equation}
S_B = a\frac{M}{1+g_M-\varepsilon} + g_S H_S({T_{\text{Suppressed}}})\,.
\label{eq:7}
\end{equation}
Similarly, at the point of perceptual suppression $E_{S}(T_{\text{Suppressed}}) \approx E_S(\infty) = S_S/(1+g_S-\varepsilon)$, and $E_M=0$, which when substituted into $D_M=0$, leads to:
\begin{equation}
S_S = \frac{1+g_S-\varepsilon}{a}(M - g_M H_M(T_{\text{Dominant}}))\,.
\label{eq:8}
\end{equation}

For slow contrast rates, dominance durations are relatively long and the exponential decay governing adaptation dynamics in the suppressed population approaches zero before the perceptual switch occurs (Figure \ref{figure:2}C). Breakthrough and suppression thresholds are, therefore, dominated by the constant terms in equations \ref{eq:7}-\ref{eq:8}. Indeed, for the lowest contrast rate ($0.0023 \times 10^{-2} \text{ ms}^{-1}$), the constant terms in the expressions for breakthrough and suppression thresholds ($S_B-S_S \approx M(\frac{a}{1+g_M-\varepsilon} - \frac{1+g_S-\varepsilon}{a}) = 0.097$) provide a reasonable approximation to the degree of hysteresis observed in simulation ($S_B-S_S = 0.095$). In this regime, the degree of hysteresis is small and chiefly determined by the strength of asymptotic inhibition from the dominant population. As the contrast rate increases, dominance durations decrease, and the time-dependent adaptation terms begin to play a non-negligible role (Figure \ref{figure:2}D), leading to a breakdown of the simple stationary approximation and an increase in hysteresis. To solve equations \ref{eq:7}-\ref{eq:8} for the general non-stationary case, we need an expression for dominance and suppression durations.

\section{Dominance duration and the depth of hysteresis}
Perceptual switches occur when the aggregate synaptic drive of the suppressed population pass through the rectification threshold from below and the population escapes suppression by overcoming competitive inhibition from the dominant population. We can, therefore, solve for dominance and suppression durations by substituting the exact solution for the dynamics of the stimulus (equation \ref{eq:2}; $S(t)=S_B-\gamma t$ for $E_S > E_M$, and $S(t)=S_S+\gamma t$ for $E_S < E_M$) along with asymptotic approximations to the  activity of the dominant population ($E_M(\infty), E_S(\infty)$) into the expression for aggregate synaptic drive (equation \ref{eq:5}) yielding:
\begin{align}
D_S &= S_S + \gamma t - a \frac{M}{1+g_M-\varepsilon} - g_S \frac{S_S}{1+g_S-\varepsilon}e^{-t/\tau_H} = 0\,, \label{eq:9} \\ 
D_M &= M - a \frac{S_B-\gamma t}{1+g_S-\varepsilon} - g_M \frac{M}{1+g_M-\varepsilon}e^{-t/\tau_H} = 0\,, \nonumber
\end{align}
where we have taken advantage of the fact that when a population is suppressed there is no recurrent activity ($\varepsilon E_{M/S}=0$) and adaptation undergoes simple exponential decay with initial conditions given by the asymptotic activity of the now suppressed population (i.e. $H_{M/S}(t) = E_{M/S}(\infty)e^{-t/\tau_H}$). $S_B$ and $S_S$ are the values of the stimulus at points of breakthrough and suppression, which serve as initial conditions for the piecewise ODE governing the dynamics of the stimulus (equation \ref{eq:2}). In essence, then, we can think of each successive period of dominance and suppression as setting the initial conditions for the following period.

Although we cannot solve for $t$ in terms of elementary functions, both expressions can be shown to have exact solutions written in terms of the Lambert W (i.e. product log) function:
\begin{align}
T_{\text{Suppressed}} &= \underbrace{\frac{1}{\gamma}\left(\frac{aM}{1+g_M-\varepsilon} - S_S\right)}_{\text{Stationary drive}} + \underbrace{\tau_H W_0\left(\frac{g_S}{\gamma\tau_H}\frac{S_S}{1+g_S-\varepsilon} e^{-\frac{1}{\gamma\tau_H}\left(\frac{aM}{1+g_M-\varepsilon}-S_S\right)}\right)}_{\text{Time-dependent drive}}, \label{eq:10} \\
T_{\text{Dominant}} &= \underbrace{\frac{1}{\gamma}\left(S_B - \frac{(1+g_S-\varepsilon)}{a}M\right)}_{\text{Stationary drive}} + \underbrace{\tau_H W_0\left(\frac{g_M M (1+g_S-\varepsilon)}{a \gamma\tau_H(1+g_M-\varepsilon)} e^{-\frac{1}{\gamma\tau_H}\left(S_B - \frac{(1+g_S-\varepsilon)}{a}M\right)}\right)}_{\text{Time-dependent drive}}\,, \nonumber
\end{align}
where $W_0$ is the principle branch of the Lambert W function. Both equations provide an excellent fit to dominance and suppression durations derived from numerical simulations across the full range of contrast rates (Figure \ref{figure:3}A). Full derivation of equation \ref{eq:10} is supplied in Appendix 2

The first term in both expressions captures the stationary contribution of the difference between the excitatory drive entering the suppressed population and the level of asymptotic inhibition from the dominant population. In fact, if we neglect the time-dependent effects of adaptation, we can derive the stationary component of equation \ref{eq:10} by substituting the exact solution for the stimulus dynamics into equations \ref{eq:7}-\ref{eq:8} and solving for $t$. This term is a monotonically decreasing function of contrast rate proportional to $\gamma^{-1}$. Notably, in isolation, because the term does not capture any time-dependent effects, including those associated with adaptation, it systematically underestimates the durations of dominance and suppression.

The second term in both expressions captures the time-dependent effects of adaptation. To intuitively understand the role of contrast rate in governing the durations of dominance and suppression (equation \ref{eq:10}), we need to take into account two key facts. First, for input $>$ 0, the Lambert W function is monotonically increasing. Second, the argument of the Lambert W function has the form $\frac{c_1}{\gamma} e^{-\frac{c_2}{\gamma} }$ where $c_1$ and $c_2$ are real-valued constants, which for all parameters used in the simulation ($\gamma << 1$), is also a monotonically increasing function of contrast rate. This term captures the fact that as contrast rates increase, there is less time for adaptation to recover, increasing the breakthrough threshold and decreasing the suppression threshold, meaning that it takes longer for input to travel between the breakthrough and suppression values. The closed-form expressions for dominance and suppression duration can, therefore, be seen as a solution to the trade-off between the opposing effects of contrast rate on dominance and suppression duration. As contrast rate increases, it takes less time for the stimulus to change from the breakthrough value to the suppression value (and vice versa), but by the same token, because adaptation has an additive effect on the breakthrough threshold, and a subtractive effect on the suppression threshold (see equations \ref{eq:7}-\ref{eq:8}), as dominance durations decrease, the distance between breakthrough and suppression values increases.

\begin{figure}[h!]
    \centering
    \includegraphics[width=.85\textwidth]{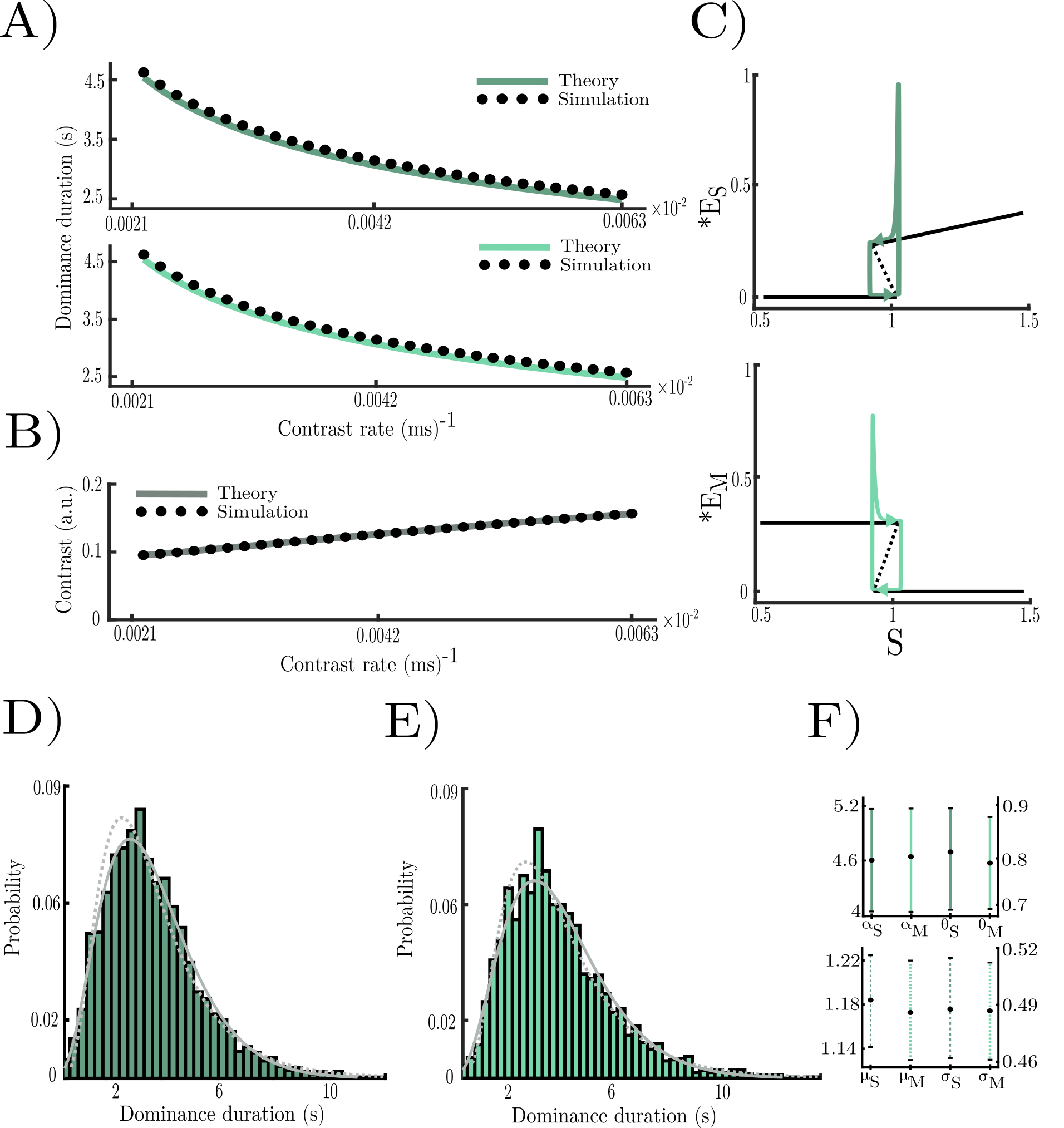}
    \caption{\footnotesize{\textbf{A)} Simulated dominance (dark green) and suppression (light green) durations across the full range of contrast rates. Numerical simulation results are shown in black dots, and values predicted from analytic theory are shown in solid lines. \textbf{B)} Simulated hysteresis depth as a function of contrast rates. Numerical simulation results are shown in black dots, and values predicted from analytic theory are shown with solid lines. \textbf{C)} Approximate bifurcation diagrams with example model trajectories for the population driven by the target stimulus (upper dark green), and the population driven by the mask (lower light green). \textbf{D-E)} Distribution of dominance (dark green) and suppression (light green) durations aggregated across all three experiments from Alais and colleagues \cite{Alais2024}. Grey lines show the fit of lognormal (dashed) and gamma (solid) distributions fit via maximum likelihood estimation. \textbf{F)} Parameters fit with 95\% confidence intervals for dominance and suppression duration distributions for gamma (upper) and lognormal (lower) distributions. Left y-axes shows $\alpha$ and $\mu$ values right y-axes show $\theta$ and $\sigma$ values.}}
    \label{figure:3}
\end{figure}

Closed-form expressions for dominance and suppression duration in hand, we can now derive an expression for hysteresis depth as a function of contrast rate. Specifically, the depth of hysteresis is given by contrast rate and the length of the dominance/suppression period. This leads to the following two equations for the depth of hysteresis between successive periods of dominance and suppression:
\begin{align}
\mathcal{H}_{S \to B} &= \gamma T_{\text{Suppressed}} \label{eq:11} \\
&= \underbrace{\left(\frac{aM}{1+g_M-\varepsilon} - S_S\right)}_{\text{Stationary drive}} + \underbrace{\gamma \tau_H W_0\left(\frac{g_S}{\gamma\tau_H}\frac{S_S}{1+g_S-\varepsilon} e^{-\frac{1}{\gamma\tau_H}\left(\frac{aM}{1+g_M-\varepsilon}-S_S\right)}\right)}_{\text{Time-dependent drive}} \nonumber \\
\mathcal{H}_{B \to S} &= \gamma T_{\text{Dominant}} \nonumber \\
&= \underbrace{\left(S_B - \frac{(1+g_S-\varepsilon)}{a}M\right)}_{\text{Stationary drive}} + \underbrace{\gamma \tau_H W_0\left(\frac{g_M M (1+g_S-\varepsilon)}{a \gamma\tau_H(1+g_M-\varepsilon)} e^{-\frac{1}{\gamma\tau_H}\left(S_B - \frac{(1+g_S-\varepsilon)}{a}M\right)}\right)}_{\text{Time-dependent drive}}\,, \nonumber
\end{align}
where $\mathcal{H}_{S \to B}$ and $\mathcal{H}_{B \to S}$ denote the hysteretic difference between successive contrast thresholds between points of suppression and breakthrough and vice versa. Crucially, once the model has converged to an equilibrium, the expressions become interchangeable (i.e. $\mathcal{H}_{S \to B} = \mathcal{H}_{B \to S}$) and provide an excellent fit to the hysteresis depth values derived from numerical simulation (Figure \ref{figure:3}B). One can understand the symmetry between periods of dominance and suppression by noting that:
\begin{align}
S_{S_{n+1}} &= S_{B_n} - \gamma T_{\text{Dominant}_n}, \label{eq:12} \\ 
S_{B_{n+1}} &= S_{S_{n+1}} + \gamma T_{\text{Suppressed}_{n+1}}\,,\nonumber 
\end{align}
which, because of the successive dependence of breakthrough and suppression thresholds, leads to,
\begin{align*}
S_{S_{n+1}} &= S_{S_n} + \gamma T_{\text{Suppressed}_n} - \gamma T_{\text{Dominant}_n}\,, \nonumber \\
S_{B_{n+1}} &= S_{B_n} - \gamma T_{\text{Dominant}_n} + \gamma T_{\text{Suppressed}_{n+1}}\,, \nonumber \\
\end{align*}
implying that once the model has converged to equilibrium, where $S_{B_{n+1}}=S_{B_n}$ and $S_{S_{n+1}}=S_{S_n}$, it must be the case that $T_{\text{Dominant}_n} = T_{\text{Suppressed}_n}$ (Figure \ref{figure:3}A-B). We can understand this symmetry geometrically by treating each perceptual switch from dominance to suppression (and vice versa) as a bifurcation and constructing bifurcation diagrams for each population with bifurcation points given by $S_B$ and $S_S$ (Figure \ref{figure:3}C). Because the distance between $S_B$ and $S_S$ is constant across periods of dominance and suppression (once the model has reached equilibrium), the time it takes for the stimulus to decrease from $S_B$ to $S_S$ is identical to the time it takes for the stimulus to increase from $S_S$ to $S_B$.

\section{Confirmation of novel prediction}

Crucially, the symmetry of dominance and suppression durations is a prediction that we can test in the empirical data of Alais and colleagues \cite{Alais2024}. If we assume that the empirical data are generated by a noisy version of an equivalent dynamical system, the distribution of dominance durations should be approximately equal to the distribution of suppression durations. To test this prediction, we took a two-pronged approach. First, we aggregated the suppression and dominance duration data across all three experiments from Alais and colleagues \cite{Alais2024}. We excluded trials with dominance or suppression durations longer than the time it takes for the stimulus to move from the minimum to the maximum allowed contrast value, as well as individual periods of dominance or suppression shorter than 500ms, as these instances likely reflect either a noise-driven perceptual switch or a failure to follow task instructions. In addition, as the prediction applies to systems close to their equilibrium value, we excluded the first two periods of dominance and suppression after trial onset. We then used a two-sample Kolmogorov-Smirnov test against the null hypothesis that the data come from the same underlying distribution. The test failed to reject the null hypothesis at the 5\% significance level ($p=0.5555$).

Second, having failed to reject the null hypothesis that the data come from the same underlying distribution, we next sought to find confirmatory evidence for the equivalence of the dominance and suppression distributions. We therefore fit gamma and lognormal distributions, which have both previously been shown to provide a good fit to bistable perception data \cite{Fox1967, Lehky1997, Levelt1967, Brascamp2005, Shpiro2007, Pastukhov2013}, to the dominance and suppression duration data via maximum likelihood estimation. Both distributions provided an excellent fit to the data (Figure \ref{figure:3}D-F) with the gamma distribution performing marginally better than the lognormal distribution in terms of the negative log likelihoods (dominant $l_{\text{gamma}}/l_{\text{lognormal}} =0.9945$, suppressed $l_{\text{gamma}}/l_{\text{lognormal}} =0.9971$)\cite[but, also see,][]{Brascamp2005}. In line with the prediction from the model, parameter estimates for dominance and suppression distributions were near identical for both the gamma ($\alpha_{\text{Dom}}= 4.64, \theta_{\text{Dom}}=0.784$; $\alpha_{\text{Sup}}=4.63, \theta_{\text{Sup}}=0.78$) and lognormal ($\mu_{\text{Dom}}=1.18, \sigma_{\text{Dom}}=0.49$; $\mu_{\text{Sup}}=1.17, \sigma_{\text{Sup}}=0.48$) distributions with almost entirely overlapping 95\% confidence intervals (Figure \ref{figure:3}F).

\section{Additional predictions}

\begin{figure}[H]
    \centering
    \includegraphics[width=0.85\textwidth]{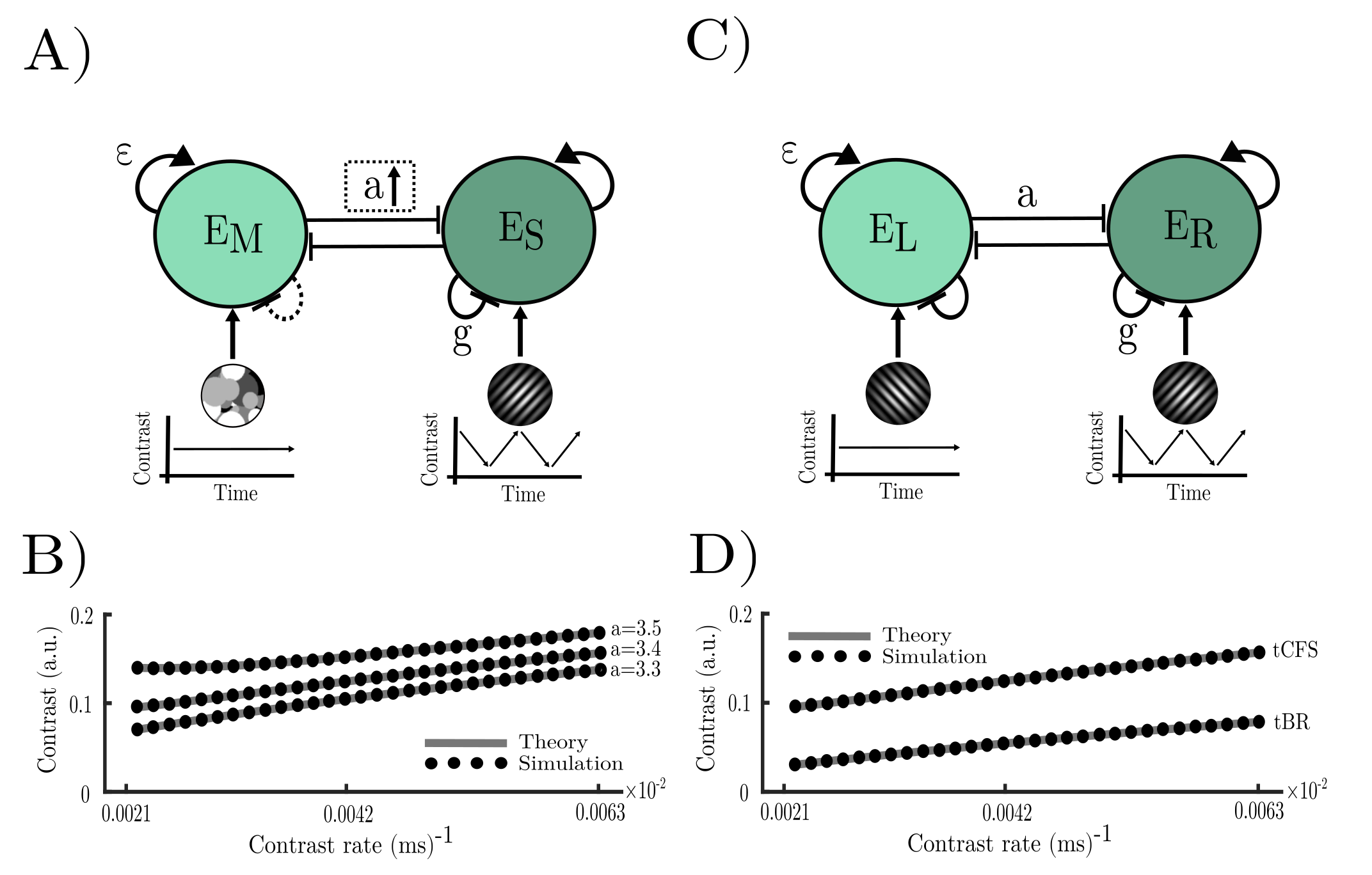}
    \caption{\footnotesize{\textbf{A)} tCFS model architecture highlighting the role of competitive inhibition (a). \textbf{B)} Hysteresis depth for tCFS simulations across the full range of contrast rates with competitive inhibition values of 3.3, 3.4, and 3.5. \textbf{C)} Model architecture for ‘tracking’ binocular rivalry (tBR). \textbf{D)} Hysteresis depth for tCFS and tBR simulations across the full range of contrast rates.}}
    \label{figure:4}
\end{figure}

Having shown that the model accurately reproduces existing data and tested a novel prediction in existing data, we next sought to move beyond the current tCFS paradigm by generating two additional predictions that can be tested in future experiments. Inspired by previous bistable perception experiments that pharmacologically manipulate baseline levels of inhibition by giving participants a GABA agonist \cite{vanLoon2013}, we simulated the same tCFS paradigm whilst simultaneously sweeping the degree of competitive inhibition ($a$) between values of 3.3-3.5 (Figure \ref{figure:4}A). We stayed within this narrow range of values as the model is highly sensitive to changes in competitive inhibition and values outside this range lead to empirically unrealistic behaviour. In line with the predictions of equation \ref{eq:10}, increasing competitive inhibition increased the depth of hysteresis across the full range of contrast rates (Figure \ref{figure:4}B), an effect that is chiefly mediated by increasing the stationary effect of inhibition. Averaged across $\mathcal{H}_{S \to B}$ and $\mathcal{H}_{B\to S}$, the relative contribution of each term to the total degree of hysteresis changed from stationary = 54.2\%, time-dependent = 45.8\% for $a=3.3$ to, stationary = 89.61\%, time-dependent = 10.39\% for $a=3.5$.

Next, we took advantage of the fact that manipulating the stimuli entering each eye provides an indirect means to manipulate the net strength of adaptation (Figure \ref{figure:4}C). Specifically, instead of driving one eye with a flashing mask (as is done in standard tCFS ), which periodically drives unadapted neurons, reducing the overall effect of adaptation \cite{Shimaoka2011}, we drive both eyes with standard grating stimuli, leaving the contrast of one eye fixed and dynamically ramping the contrast of the other. This equalises the adaptation strength of each population. We refer to this extension of the tCFS paradigm as tracking binocular rivalry (tBR). We simulated tBR in the model by setting $g_M = g_S$. tBR almost halves the depth of hysteresis seen in standard tCFS (Figure \ref{figure:4}D), an effect predicted by equation \ref{eq:10} that is primarily mediated by a reduction in the stationary effect of inhibition due to the increased accumulation of adaptation in the dominant population, reducing the asymptotic firing rate of the population driven by the fixed contrast stimulus. This reduces the breakthrough threshold when the dynamic stimulus-driven population is suppressed, which in turn reduces the stationary effect of inhibition when the same population is dominant, leading to a net decrease in the depth of hysteresis. Again, averaging across $\mathcal{H}_{S \to B}$ and $\mathcal{H}_{B \to S}$, the relative contribution of each term to the total degree of hysteresis changed from stationary = 76.8\%, time-dependent = 23.2\% for tCFS, to stationary = 39.22\%, time-dependent = 60.78\% for tBR.

\section{Discussion}

Building on our companion paper \cite{whyte2025minimal}, which generalised Wilson's \cite{Wilson2007} minimal model of binocular rivalry to tracking continuous flash suppression (tCFS), we developed a quantitative account of the mechanisms governing awareness and suppression in tCFS. Visual rivalry is one of the few experimental paradigms in psychology and neuroscience where there are law-like explanations and accompanying closed-form expressions, analogous to those observed in the physical sciences \cite[see also, ][]{Li2014}. The key contribution of this paper is the derivation of closed-form expressions for the duration of dominance and suppression periods for the type of non-stationary stimuli used in tCFS, and an accompanying expression for the depth of hysteresis (i.e., suppression depth) between breakthrough and suppression thresholds.

In addition to explaining existing data, the expression for the depth of hysteresis led to the prediction that the distributions of dominance and suppression durations should be equivalent, which we tested and confirmed, in previously collected psychophysical data from Alais and colleagues \cite{Alais2024}. Crucially, we also used the model and the accompanying expression for the depth of hysteresis to move beyond existing data and expose the model to potential falsification in two future experiments. Specifically, pharmacological manipulation of competitive inhibition should increase suppression depth, and equalising the strength of adaptation between competing neural populations by driving each population with standard grating stimuli instead of using a flashing mask should approximately halve the depth of hysteresis between breakthrough and suppression thresholds.

Lastly, the dynamical principles we employed in this study are similar to a wide range of multistability tasks. For instance, previous studies based on computational modeling \cite{Theodoni2011, Panagiotaropoulos2013} suggested that the population-level neural dynamics underlying binocular rivalry \cite{Theodoni2011} are similar to the dynamical regime underlying binocular flash suppression (BFS; \cite{Panagiotaropoulos2013}). In our companion paper \cite{whyte2025minimal}, we demonstrated that this is the case for binocular rivalry and tCFS. And \cite{Shimaoka2011} showed that the same principles that apply to binocular rivalry apply to standard CFS. Given that all such circuit models use similar dynamical principles, parameters learned from any of these paradigms (BFS, BR, CFS, or tCFS) should also inform the neural dynamics and switching behavior of others. Thus, we believe, the neurodynamical implications of this theoretical study will likely go beyond tCFS, and give insight into a wide range of rivalry-like paradigms. 

In sum, the model presented in this paper extends insights from quantitative models of binocular rivalry \cite{Wilson2007}, and CFS \cite{Shimaoka2011}, to tCFS. In tCFS, unlike binocular rivalry or standard CFS, it is possible to measure of the threshold for awareness and suppression, and depth of perceptual hysteresis separating the two thresholds, an effect we explain quantitatively in terms of the trade-off between the (relatively) stationary effects of inhibition and external drive, and the time-dependent effects of adaptation.

\section*{Data and code availability}
Code to reproduce all simulations can be downloaded from: \hyperlink{https://github.com/cjwhyte/tCFS}{https://github.com/cjwhyte/tCFS}. The empirical data from \cite{Alais2024} analysed in the paper can be downloaded from the open science framework: \hyperlink{https://osf.io/nzp9v/}{https://osf.io/nzp9v/}.

\clearpage

\section*{Appendix}
\subsection*{Appendix 1: model parameters}

All simulations were run by integrating equation \ref{eq:1} numerically with the parameters in table 1 using the forward Euler method with a time step of $dt=0.1$ ms in MATLAB 2023b.

\begin{table}[h!]
\centering
\begin{tabular}{p{2cm} p{7cm} p{4cm} p{1cm}}
\hline
\textbf{Parameter} & \textbf{Description} & \textbf{Value}  & \textbf{Units}\\
\hline
$\tau_E$ & Neuronal population time constant & 15  & ms\\
$\tau_H$ & Hyperpolarizing adaptation current time constant & 1000 & ms\\
$M$ & Constant drive of mask & 0.8 & a.u.\\
$S$ & Initial condition for stimulus drive & $S_{t_0}=1.2$ & a.u.\\
$\varepsilon$ & Strength of excitatory recurrent projections & 0.05 & a.u.\\
$a$ & Strength of competitive inhibition & 3.3-3.5 (base = 3.4) & a.u.\\
$g_M$ & Strength of mask population adaptation current & 1.7  & a.u.\\
$g_S$ & Strength of stimulus population adaptation current & 3 & a.u.\\
$\gamma$ & Rate of contrast change & ($0.0021 - 0.0063) \times 10^{-2}$ & a.u.\\
\hline
\end{tabular}
\caption{\footnotesize{Parameter description, values, and units of the model described by equations 1-6 for tCFS simulations.}}
\end{table}

\subsection*{Appendix 2: derivation of dominance and suppression duration formula}
Although it is not trivial to solve for $t$ in equation \ref{eq:9}, based on the assumptions and approximations described in the main text, both equations can be shown to have the form $y=xe^x$ which for real $x$ and $y$, on the restricted domain $x \ge 0$, have an inverse function $W_0$ (i.e. $x=W_0(xe^x)$), where $W_0$ is the principle branch of the Lambert W function.

Starting with line one of equation \ref{eq:9}, where for compactness of notation we let $\bar{E}_M = E_M(\infty) = M/(1+g_M-\varepsilon)$ and $\bar{E}_S = E_S(\infty) = S_S/(1+g_S-\varepsilon)$, we have:
\begin{align*}
S_S + \gamma t - a\bar{E}_M - g_S \bar{E}_S e^{-t/\tau_H} &= 0 \\
a\bar{E}_M - S_S - \gamma t &= -g_S \bar{E}_S e^{-t/\tau_H} \\
(a\bar{E}_M - S_S - \gamma t) e^{t/\tau_H} &= -g_S \bar{E}_S \\
\left(t + \frac{S_S - a\bar{E}_M}{\gamma}\right) e^{t/\tau_H} &= \frac{g_S \bar{E}_S}{\gamma} \\
\left(\frac{t}{\tau_H} + \frac{S_S - a\bar{E}_M}{\gamma\tau_H}\right) e^{\frac{t}{\tau_H} + \frac{S_S - a\bar{E}_M}{\gamma\tau_H}} &= \frac{g_S \bar{E}_S}{\gamma\tau_H} e^{\frac{S_S - a\bar{E}_M}{\gamma\tau_H}} \\
\frac{t}{\tau_H} + \frac{S_S - a\bar{E}_M}{\gamma\tau_H} &= W_0\left(\frac{g_S \bar{E}_S}{\gamma\tau_H} e^{\frac{S_S - a\bar{E}_M}{\gamma\tau_H}}\right)\,.
\end{align*}
After minor rearrangement and expanding both $\bar{E}_M$ and $\bar{E}_S$ we obtain,
\begin{align*}
T_{\text{Suppression}} = \underbrace{\left(\frac{aM}{1+g_M-\varepsilon} - S_S\right)}_{\text{Stationary drive}} + \underbrace{\gamma \tau_H W_0\left(\frac{g_S}{\gamma\tau_H}\frac{S_S}{1+g_S-\varepsilon} e^{-\frac{1}{\gamma\tau_H}\left(\frac{aM}{1+g_M-\varepsilon}-S_S\right)}\right)}_{\text{Time-dependent drive}}\,.
\end{align*}

The derivation of the dominance time from line two of equation \ref{eq:9} is slightly more complex, as changes in the strength of the stimulus have a delayed effect on the subthreshold dynamics of the mask-driven population. This is because the strength of asymptotic inhibition from the dominant population is dependent on adaptation. To compensate for the delay, we added a correction term to the expression for asymptotic firing rate $\tilde{S}_B = S_B - \gamma\tau_{\text{Delay}}$. We derived the value of the correction term by finding the value of $\tau_{\text{Delay}}$ which best minimised the difference between the closed-form expression for the asymptotic firing rate at suppression points ($E_S(\infty) = (\tilde{S}_B-\gamma t)/(1+g_S-\varepsilon)$) and the value derived from simulation. We averaged this value across contrast rates, giving $\tau_{\text{Delay}}=760$ ms for the tCFS simulations, and $\tau_{\text{Delay}}=634$ ms for the tBR simulations.

Correction in hand we arrive at,
\begin{align*}
M - \frac{a(\tilde{S}_B - \gamma t)}{1+g_S-\varepsilon} - g_M \bar{E}_M e^{-t/\tau_H} &= 0 \\
M - \frac{a \tilde{S}_B}{1+g_S-\varepsilon} + \frac{a \gamma t}{1+g_S-\varepsilon} &= g_M \bar{E}_M e^{-t/\tau_H}\\
\left(M - \frac{a \tilde{S}_B}{1+g_S-\varepsilon} + \frac{a \gamma t}{1+g_S-\varepsilon}\right) e^{t/\tau_H} &= g_M \bar{E}_M\\
\left(\frac{t}{\tau_H} - \frac{a \tilde{S}_B - M(1+g_S-\varepsilon)}{a \gamma \tau_H}\right)e^{\frac{t}{\tau_H} - \frac{a \tilde{S}_B - M(1+g_S-\varepsilon)}{a \gamma \tau_H}} &= \frac{g_M \bar{E}_M(1+g_S-\varepsilon)}{a \gamma \tau_H}e^{-\frac{a \tilde{S}_B - M(1+g_S-\varepsilon)}{a \gamma \tau_H}}\\
\frac{t}{\tau_H} - \frac{a \tilde{S}_B - M(1+g_S-\varepsilon)}{a \gamma \tau_H} &= W_0\left( \frac{g_M \bar{E}_M(1+g_S-\varepsilon)}{a \gamma \tau_H}e^{-\frac{a \tilde{S}_B - M(1+g_S-\varepsilon)}{a \gamma \tau_H}}\right)\,,
\end{align*}
expanding $\bar{E}_M$ and solving for $t$, we obtain,
\begin{align*}
T_{\text{Dominance}} = \underbrace{\frac{1}{\gamma}\left(\tilde{S} - \frac{(1+g_S-\varepsilon)}{a}M\right)}_{\text{Stationary drive}} + \underbrace{\tau_H W_0\left(\frac{g_M M (1+g_S-\varepsilon)}{a \gamma\tau_H(1+g_M-\varepsilon)} e^{-\frac{1}{\gamma\tau_H}\left(\tilde{S} - \frac{(1+g_S-\varepsilon)}{a}M\right)}\right)}_{\text{Time-dependent drive}}\,.
\end{align*}
Each expression is dependent on the stimulus values at suppression ($S_S$) and breakthrough points ($S_B$), respectively. Instead of simply pulling these values from the simulation, we leveraged the successive dependence of breakthrough and suppression values, and the duration of suppression and dominance, to write equation \ref{eq:10} and equation \ref{eq:12} as a set of difference equations which we iterated until convergence (see, Algorithm \ref{algo:domTimeUpdate}).
\begin{algorithm}[H]
\caption{Iterative update of $S_B$, $T_{\text{Dominance}}$, and $T_{\text{Suppression}}$}
\begin{algorithmic}[1]
    \State Initialise: $S_B \gets S_{t_0}$
    \State Initialise: $T_{\text{Dominance}} \gets \dfrac{1}{\gamma}\left(S_B - \dfrac{(1+g_S-\varepsilon)}{a}M\right)$
    \Repeat
        \State $S_S \gets S_B - \gamma T_{\text{Dominance}}$
        \State $T_{\text{Suppression}} \gets \dfrac{1}{\gamma}\left(\dfrac{aM}{1+g_M-\varepsilon} - S_S\right)$
        \Statex \hspace{3.5em} $+ \tau_H W_0\!\left(\dfrac{g_S}{\gamma\tau_H}\dfrac{S_S}{1+g_S-\varepsilon} 
        e^{-\tfrac{1}{\gamma\tau_H}\left(\tfrac{aM}{1+g_M-\varepsilon}-S_S\right)}\right)$
        \State $S_B \gets S_S + \gamma T_{\text{Suppression}}$
        \State $T_{\text{Dominance}} \gets \dfrac{1}{\gamma}\left(\tilde{S}_B - \dfrac{(1+g_S-\varepsilon)}{a}M\right)$
        \Statex \hspace{4.5em} $+ \tau_H W_0\!\left(\dfrac{g_M M (1+g_S-\varepsilon)}{a \gamma\tau_H(1+g_M-\varepsilon)} 
        e^{-\tfrac{1}{\gamma\tau_H}\left(\tilde{S}_B - \tfrac{(1+g_S-\varepsilon)}{a}M\right)}\right)$
    \Until{convergence}.
\end{algorithmic}
\label{algo:domTimeUpdate}
\end{algorithm}
We initialised the breakthrough value of the stimulus with the initial condition of the stimulus used in the simulations, and the initial dominance duration value with the stationary component of equation \ref{eq:10}. The difference equations rapidly converged across all contrast rates to the equilibrium values shown in Figure \ref{figure:3} and Figure \ref{figure:4}. This approach allowed us to arrive at closed-form expressions that predicted the behaviour of the model without the use of any information derived from the simulation results themselves (with the exception of the delay correction).

\clearpage

\printbibliography

\end{document}